\documentclass[twocolumn,10pt]{article}

\usepackage[utf8]{inputenc}
\usepackage[margin=2cm]{geometry}
\usepackage{amsmath,amssymb}
\usepackage{graphicx}
\usepackage{booktabs}
\usepackage{cite}
\usepackage{hyperref}
\usepackage{abstract}
\usepackage{titlesec}
\usepackage{enumitem}
\usepackage{array}
\usepackage{longtable}
\usepackage{tabularx}
\usepackage{booktabs}
\usepackage[table]{xcolor}
\usepackage{enumitem}
\usepackage{tabularx, booktabs, xcolor, amsmath}
\usepackage[table]{xcolor}
\definecolor{rowgrey}{gray}{0.95}
\titleformat{\section}{\normalfont\Large\bfseries}{\thesection}{1em}{}
\titleformat{\subsection}{\normalfont\large\bfseries}{\thesubsection}{1em}{}
\titleformat{\subsubsection}{\normalfont\normalsize\bfseries}{\thesubsubsection}{1em}{}

\title{\vspace{-1cm}\textbf{Quantum Annealing for Combinatorial Optimization: Foundations, Architectures, Benchmarks, and Emerging Directions}}

\author{
    \textbf{Rudraksh Sharma}\textsuperscript{1,*}, 
    \textbf{Ravi Katukam}\textsuperscript{1}, 
    \textbf{Arjun Nagulapally}\textsuperscript{1} \\
    \textsuperscript{1}\textit{AIONOS, Financial District, Hyderabad, Telangana, India} \\
    \texttt{\ Rudraksh.sharma@aionos.ai, katukam.Ravi@aionos.ai, Arjun.Nagulapally@aionos.ai} \\
    \textsuperscript{*}Corresponding author: \texttt{Rudraksh.sharma@aionos.ai}
}

\date{}

\begin{document}
\twocolumn[
\begin{@twocolumnfalse}
\maketitle

\begin{abstract}
Critical decision-making issues in science, engineering, and industry are based on combinatorial optimization; however, its application is inherently limited by the NP-hard nature of the problem. A specialized paradigm of analogue quantum computing, quantum annealing (QA), has been proposed to solve these problems by encoding optimization problems into physical energy landscapes and solving them by quantum tunnelling systematically through exploration of solution space. This is a critical review that summarizes the current applications of quantum annealing to combinatorial optimization and includes a theoretical background, hardware designs, algorithm implementation strategies, encoding and embedding schemes, protocols to benchmark quantum annealing, areas of implementation, and links with the quantum algorithms implementation with gate-based hardware and classical solvers. We develop a unified framework, relating adiabatic quantum dynamics, Ising and QUBO models, stoquastic and non-stoquastic Hamiltonians, and diabatic transitions to modern flux-qubit annealers (Chimera, Pegasus, Zephyr topologies), and emergent architectures (Lechner-Hauke-Zoller systems, Rydberg atom platforms), and hybrids of quantum and classical computation. Through our analysis, we find that overhead in embedding and encoding is the largest determinant of the scalability and performance (this is not just the number of qubits). Minor embeddings also usually have a physical qubit count per logical variable of between 5 and 12 qubits, which limits effective problem capacity by 80-92\% and, due to chain-breaking errors, compromises the quality of solutions. We show that quantum annealing can offer empirical utility that can be measured quantitatively as a hybrid refinement method and not a solver by systematic study of applications across the field of transportation logistics, optimization of energy systems, robotics, finance, molecular design, and machine learning. We critically analyse the current types of benchmarking, noting such methodological shortcomings as a lack of selective reporting of results, poor classical baselines and the exclusion of preprocessing overhead to distort real performance comparisons. In addition, we examine the structural connection between quantum annealing and gate-based variational algorithms (QAOA, VQE) on what happens to them in the limit of fine-grained control and what basic restrictions are dictated by stoquastic constraints. We finish by establishing priority directions in research: comprehensively define annealing hardness problems regardless of size, develop non-stoquastic control, devise useful embedding algorithms that are performance guaranteed, develop standardized benchmarking protocols, and have principled measures of quantum advantage. The review is a perfect guide for everyone conducting research and practitioners in the field of quantum-enhanced combinatorial optimization.
\end{abstract}
\vspace{0.5cm}
\end{@twocolumnfalse}
]

\section{Introduction}

\subsection{The Combinatorial Optimization Challenge}
Decision-making in science, engineering and industry: Combinatorial optimization problems is the problem of choosing the best configurations in exponentially large discrete solution spaces. Examples of these include scheduling manufacturing processes, picking paths of delivery trucks on routes, creating molecular compounds with the properties required and assigning financial or portfolio combinations of investments to maximise risk-adjusted returns. These issues are defined by the fact that they scale exponentially: as N binary decision variables are added to a system, it grows to $2^{N}$ choices of configuration. Even relatively low-order examples of 50 scheduling tasks, 20 delivery sites or 100 portfolio assets offer solution spaces of higher than $10^{15}$ solutions, making exhaustive enumeration computationally infeasible.
Many combinatorial problems are intractable, not only empirically, but also structurally defined, using computational complexity theory. There are a significant number of problems of practical importance (NP-hard problems) whose exact algorithms are not known in practice, even after many years of research. This inherent hardness can be occasioned by the combination of constraints and dependency to form a combinatorial explosion, and not necessarily modelling complexity. The Quadratic Unconstrained Binary Optimisation (QUBO) framework offers a coherent mathematical abstraction of the representation of these problems, and by its formal coincidence with the Ising spin glass model, gives a direct physical understanding that provides quantum annealing strategies 
 \cite{chowla1950,glover2019,cipra1987}.

\subsection{Classical Approaches and their Limitations}

There are three main strategies that are used by classical optimization algorithms to solve combinatorial problems. Precise algorithms like branch-and-bound and mixed-integer linear programming are guaranteed to provide optimal solutions but have worst-case time advantages of an exponential nature. Approximation algorithms offer solutions of the form that can answer problems in time polynomial and whose performance can be proved (however, with restrictions to particular problem structure). Metaheuristics such as simulated annealing, tabu search, genetic algorithms, and particle swarm optimization do not provide any guarantees of optimality in favor of scalability.
As a classical analogue of quantum annealing, much attention should be paid to simulated annealing. It uses thermodynamic variations controlled by a Boltzmann distribution to get out of local minima, and the temperature parameter is reduced by a cooling schedule. Logarithmic cooling rates are needed to theoretically converge to global optima, which cannot be practiced in practice because of their sluggishness. Although metaheuristics have demonstrated impressive performance on several problem instances due to decades of algorithmic optimization, the concepts are eventually limited to basic cases when they are run on adversarial constructed instances or problems with exponentially many local minima separated by high energy barriers. \cite{kadowaki1998,hauke2020}.

\subsection{Quantum Annealing: Principles and Evolution}

Quantum annealing was a physics-inspired optical model that began to appear in the late 1990s, based upon quantum mechanical effects, quantum tunnelling using energy barriers, to potentially speed up combinatorial search. Its main concept is to encode an optimisation problem as a problem Hamiltonian $H_{\text{prob}}$ in which the ground state is the optimal solution to this problem, initialise a quantum system in the easily prepared ground state of a driver Hamiltonian $H_{\text{driver}}$ and slowly drive an interpolation between the Hamiltonians, based on a time-dependent time schedule. When this evolution is carried out slowly (adiabatically), the quantum adiabatic theorem would ensure that the system is in its adiabatic ground state, and would finally give the best solution to the problem.\cite{kadowaki1998,morita2008,peruzzo2014,farhi2014}.
The important difference between the classical simulated annealing and the one discussed here is the exploration mechanism. Compared to simulated annealing, which uses thermal activation to overcome energy barriers (probability decreases exponentially with barrier height), quantum annealing takes advantage of quantum tunnelling to overcome barriers (probability decreases exponentially with barrier width). When the problem geometry has narrow, tall barriers, quantum tunnelling may give exponential benefit as compared to thermal hopping. On the other hand, thermal methods can be more effective when the variety of problems has broad barriers, or entropic bottlenecks, and it is clear that the performance of quantum annealing is problem-dependent.
In theoretical studies, quantum annealing was originally defined as a form of adiabatic quantum computation (AQC), allowing it to be studied as a spectral gap dynamics, many-body quantum physics, and computational complexity theory. Follow-up commercialisation of programmable quantum annealing devices by D-Wave Systems started with a 128-qubit device in 2011 and expanded up to more than 5,600 qubits by 2024, triggering expansive empirical studies in the areas of application. This development has resulted in a rich ecosystem that includes theoretical study, numerical modelling, and experimental implementation.

\begin{table}[htbp]
\centering
\caption{Mapping Optimization Concepts to Quantum Annealing}
\label{tab:qa_mapping}
\footnotesize 
\renewcommand{\arraystretch}{1.4} 
\definecolor{tablegrey}{gray}{0.9} 

\begin{tabularx}{\columnwidth}{@{} >{\raggedright\arraybackslash}p{1.7cm} >{\raggedright\arraybackslash}X >{\raggedright\arraybackslash}X >{\raggedright\arraybackslash}X @{}}
\toprule
\textbf{Optimization Concept} & \textbf{QA / Physics Analog} & \textbf{Mathematical Object} & \textbf{Practical Implication} \\ \midrule
\rowcolor{tablegrey} Objective function & Problem Hamiltonian & Ising / QUBO & Defines energy landscape \\ 
Feasible solution & Ground state & Lowest eigenstate & Optimal solution \\ 
\rowcolor{tablegrey} Constraint violation & Energy penalty & Coupling strength & Precision bottleneck \\ 
Local minimum & Excited state & Local eigenstate & Trapping risk \\ 
\rowcolor{tablegrey} Neighborhood move & Quantum tunneling & Transverse field & Barrier traversal \\ \bottomrule
\end{tabularx}
\end{table}

\subsection{The Reality of Quantum Annealing Performance}

Despite hardware scaling and algorithmic refinements, the anticipated exponential quantum speedups have remained largely confined to carefully constructed benchmark problems. Rigorous comparative studies reveal that quantum annealers can match or occasionally exceed specialized classical heuristics on problem families exhibiting specific structural properties—particularly those with tall, narrow energy barriers conducive to quantum tunneling. However, for the majority of real-world applications involving complex constraint structures, practical deployments invariably employ hybrid quantum-classical workflows where quantum annealing serves as a refinement subroutine following substantial classical preprocessing.
A central thesis of this review is that the primary barrier to scalable quantum advantage is not quantum decoherence, limited coherence times, or insufficient qubit counts, but rather the classical computational overhead associated with problem encoding and minor embedding. Current quantum annealing hardware implements sparse connectivity topologies (6-15 physical connections per qubit), while real-world optimization problems typically require dense or complete logical connectivity. Minor embedding—the process of mapping logical problem variables onto chains of ferromagnetically coupled physical qubits—introduces substantial overhead, typically requiring 5-12 physical qubits per logical variable and reducing effective problem capacity by 80-92\% \cite{glaetzle2017}.
Additionally, constraint-heavy problem formulations demand penalty coefficients that dominate objective terms, often requiring coefficient dynamic ranges of 100:1 to 1000:1 to ensure constraint satisfaction. However, current hardware provides coefficient precision limited to approximately 0.1-1\%, effectively constraining dynamic range to ~100:1. When encoding requirements exceed hardware precision capabilities, constraint violations occur in 30-60\% of solution samples, necessitating classical post-processing and degrading solution quality.

\subsection{Purpose and Contributions}

This review provides a comprehensive synthesis of quantum annealing for combinatorial optimization, adopting a critical perspective that evaluates both capabilities and fundamental limitations. We systematically examine theoretical foundations, hardware architectures and connectivity constraints, algorithmic strategies and control protocols, encoding and embedding methodologies, benchmarking practices and performance evaluation, application domains and empirical results, relationships to alternative quantum optimization paradigms, and current limitations with future research directions.
Our principal contributions include: a comprehensive taxonomy of quantum annealing hardware, algorithms, and encoding strategies; critical evaluation of benchmarking methodologies identifying systematic flaws in performance reporting; quantitative analysis demonstrating that embedding overhead, rather than qubit count, dominates scalability; synthesis of application domain results emphasizing the necessity of hybrid workflows; and identification of specific technical barriers with feasibility assessments for future research directions. This review is intended for quantum computing researchers seeking comprehensive understanding of quantum annealing's current state, optimization researchers evaluating quantum approaches for their problem domains, and practitioners assessing technology maturity for deployment.

\section{Mathematical and Physical Foundations}

The conceptual foundation of quantum annealing lies in encoding classical combinatorial optimization problems as quantum Hamiltonians, then leveraging quantum dynamics to evolve a prepared initial state toward the Hamiltonian's ground state, which encodes the desired optimal solution.

\subsection{Adiabatic Quantum Computation}

Quantum adiabatic computation (QAC) is based on quantum adiabatic theorem. Given Quantum adiabatic computation (QAC) is based on quantum adiabatic theorem. Given a time-dependent Hamiltonian, which depends on time, and interpolates two Hamiltonians: an initial Hamiltonian $H_{\text{init}}$ and a problem Hamiltonian, encoded in its Hamiltonian distribution, which is $H_{\text{prob}}$. TAs this interpolation is performed at a slow rate, the system will be in the instantaneous ground state at all times during the evolution process, and finally reach the ground state of $H_{\text{prob}}$.
The principle is literally realized in transverse field Ising models, in which a weakening driver Hamiltonian is used to push the system into the ground state of the problem. The minimum spectral gap at any given time is an adiabatic convergence which strictly obeys the inverse-square rule of the runtime. Less gaps require longer periods of evolution to reduce excitations, with this correlation existing in digitized as well as analogue adiabatic quantum computing paradigms. In the case of gate-based architectures, the continuum-valued Hamiltonian evolution is split into discrete Trotter slices, which are implemented by quantum gates, which hence shed light on the feasibility of solving multi-qubit problems by a large batch of gates\cite{kadowaki1998,morita2008,rajak2023}.

\subsection{Quantum Annealing versus Adiabatic Quantum Computation}

While adiabatic quantum computation assumes ideal unitary evolution with perfect coherence, practical quantum annealing systems operate as open quantum systems subject to environmental interactions. Quantum annealing explicitly permits diabatic transitions and incorporates thermal effects, decoherence, and relaxation mechanisms that can paradoxically accelerate convergence by facilitating escape from shallow local minima. Consequently, quantum annealing should not be regarded merely as a noisy implementation of AQC, but rather as a distinct optimization strategy combining quantum tunneling, thermal activation, and device-specific characteristics to navigate complex energy landscapes. This distinction has important implications. Whereas AQC runtime is fundamentally limited by minimum spectral gap scaling, quantum annealing can potentially exploit diabatic transitions and thermal relaxation to overcome energy barriers more efficiently than pure adiabatic evolution. However, this also means quantum annealing performance depends critically on noise characteristics, temperature, and control precision—factors absent from idealized AQC analysis\cite{hauke2020,yulianti2022}.

\subsection{Ising and QUBO Models}

Quantum annealing solves the optimization problems by the mapping of the binary decision variables to the spin configurations. The Ising model takes the form of a classical Hamiltonian, with spins being the coded variables $s_i \in \{-1, +1\}$.

\begin{equation}
H_{\text{Ising}}(s) = \sum_{i} h_{i}s_{i} + \sum_{i < j} J_{ij}s_{i}s_{j}
\end{equation}

The QUBO (Quadratic Unconstrained Binary Optimization) problem model substitutes spins by binary variables, $x_{i} \in \{0,1\}$,], i.e. the objective is $x^T$Qx, and Q is a symmetric matrix which describes all the interactions. This is accomplished by a direct transformation via $s_{i} = 2x_{i} - 1$ to make the two mathematical formulations mathematically identical, but this makes them hardware-distinguishable.\cite{morita2008,rajak2023}.
An example traditional method of converting a real-world combinatorial problem has the following steps: First, the goal function is reformulated as an expression in discrete variables, followed by an encoding of domain choices into binary decision variables. The imposition of constraints is done using quadratic penalty terms and as much as a high-order interaction exists, it can be systematically reduced to quadratic form but at the cost of additional auxiliary variables. This is followed by the formulation of the quadratic unconstrained binary optimization (QUBO) problem that is converted to an Ising model of spins of ±1 which can be implemented on physical quantum annealing machines.

\subsection{Stoquastic vs Non-Stoquastic Hamiltonians}
The stoquastic property of Hamiltonians fundamentally determines classical simulation complexity. A Hamiltonian is stoquastic if its off-diagonal matrix elements are real and non-positive in the computational basis. Stoquastic Hamiltonians avoid the quantum Monte Carlo sign problem, enabling efficient classical simulation via path-integral Monte Carlo methods \cite{bravyi2015,marvian2019,bravyi2007,vinci2017}.

\begin{table}[htbp]
\centering
\caption{Comparison of QA Properties}
\label{tab:stoquastic_comparison}
\footnotesize
\renewcommand{\arraystretch}{1.5}
\definecolor{rowgrey}{gray}{0.95}

\rowcolors{2}{rowgrey}{white}

\begin{tabularx}{\columnwidth}{@{} >{\raggedright\arraybackslash}X l l @{}}
\toprule
\textbf{Property} & \textbf{Stoquastic QA} & \textbf{Non-Stoquastic QA} \\ \midrule
Sign problem & Absent & Present \\ 
Classical simulability & Efficient (QMC) & Hard \\ 
Typical phase transition & First order & Potentially second order \\ 
Hardware availability & Current & Experimental \\ 
Advantage potential & Limited & High (theoretical) \\ \bottomrule
\end{tabularx}
\end{table}
This distinction has profound implications for quantum advantage. Standard transverse-field Ising quantum annealers implement stoquastic Hamiltonians, meaning their dynamics can be efficiently simulated classically. Thus, any quantum advantage from stoquastic QA must arise from practical factors (constant-factor speedups, favorable scaling constants) rather than asymptotic complexity separation. Non-stoquastic Hamiltonians break this classical simulability, potentially enabling exponential quantum speedups, but require more complex hardware implementations not yet available in commercial systems.

\subsection{Diabatic Transitions and Quantum Phase Transitions}
While ideal quantum annealing assumes adiabatic evolution, practical implementations frequently exhibit diabatic dynamics where the system undergoes transitions between energy levels, particularly near quantum phase transitions where energy gaps become exponentially small. These transitions manifest in two primary forms: first-order phase transitions exhibiting exponentially vanishing gaps (in system size), and second-order transitions with power-law gap scaling. Paradoxically, carefully controlled diabatic transitions can sometimes outperform purely adiabatic protocols by exploiting thermal relaxation or designed non-adiabatic passages to escape unfavorable regions of the energy landscape. Advanced control techniques including non-linear annealing schedules, pause-relax-resume strategies, and non-stoquastic driving Hamiltonians aim to convert unfavorable first-order transitions into more tractable second-order transitions, potentially achieving substantial runtime improvements while maintaining stoquastic systems in an intermediate regime between classical NP-hardness and quantum simulability \cite{hauke2020}.

\section{Architecture and Hardware Landscape}
Quantum annealers use continuous-time dynamics
, which are defined by an Ising model, where system behavior is determined by a set of factors that include hardware connectivity, hardware noise, properties of coherence, and control param precision. Currently surveyed information includes flux-qubit devices, the Lechner-Hauke-Zoller (LHZ) system, and Rydberg-atom systems as well as up-to-date control methods, including temporal pausing and reverse annealing, and the main noise sources that limit scalability.

\subsection{Flux-Qubit Based Annealers}

Superconducting flux-qubit annealers, such as those made by D-Wave technologies, use transverse-field Ising Hamiltonians by encoding logic Ising spins via bistable Josephson-junction loops. During the last several years, the architecture of connectivity has evolved to the rather sparse Chimera networks up to the higher-density Pegasus and Zephyr (Advantage-2) networks, thus reducing the overheads of minor embedding. Simulations These systems realize logical spins by using ferromagnetically coupled systems of physical qubits and the quality of the embedding has a significant impact on system-wide performance  \cite{glaetzle2017}.

Physical qubits in these systems do not directly represent logical problem variables. Instead, logical spins are realized through chains of ferromagnetically coupled physical qubits, with chain strength calibrated to maintain coherent logical behavior despite individual qubit variations and noise. The quality and stability of these chains directly impact overall system performance, with chain breaks (logical qubits fragmenting into inconsistent physical qubit states) representing a dominant error mode.

\subsection{Connectivity and Embedding Constraints}

The sparse physical connectivity of superconducting quantum annealers necessitates minor embedding: the mapping of logical problem graphs onto the physical hardware topology. This process consumes multiple physical qubits per logical variable to simulate couplings not directly available in the hardware graph. The resulting embedding overhead severely constrains effective problem size—a 5,000 physical qubit device typically supports only 400-800 logical variables after embedding, depending on problem graph structure.
Embedding quality significantly impacts performance through multiple mechanisms: longer chains increase susceptibility to chain breaks, consume available coupling strength (reducing effective energy scale), and amplify sensitivity to control noise and temperature. Topology-aware embedding strategies are therefore essential, as mismatch between problem connectivity structure and hardware topology fundamentally limits computational efficiency. This embedding bottleneck, represents arguably the most significant barrier to practical quantum annealing scalability.\cite{glaetzle2017,raymond2020}.

\subsection{Alternative Architectures}

Lechner-Hauke-Zoller (LHZ) architecture addresses connectivity limitations through a fundamentally different approach: encoding Ising couplings into local multi-body constraints rather than pairwise physical qubit interactions. This eliminates the explicit need for minor embedding, providing all-to-all logical connectivity through local constraint satisfaction. Geometrical brachistochrone techniques can further optimize annealing paths within the LHZ framework, potentially achieving rapid convergence. Rydberg atom platforms offer another alternative, exploiting strong, tunable long-range interactions between atoms in highly excited Rydberg states [16,18]. These systems enable flexible connectivity patterns, support both analog and digital control paradigms, and can implement non-stoquastic Hamiltonians beyond the capabilities of flux-qubit systems. While less mature than superconducting platforms, Rydberg systems demonstrate promising coherence properties and control precision \cite{glaetzle2017,leitner2023,zhao2024}.

\begin{table}[htbp]
\centering
\caption{Architecture Comparison}
\label{tab:architecture_comp}
\footnotesize
\renewcommand{\arraystretch}{1.5}
\rowcolors{2}{gray!10}{white}

\begin{tabularx}{\columnwidth}{@{} l X l l l @{}}
\toprule
\textbf{Architecture} & \textbf{Connectivity} & \textbf{Control} & \textbf{Noise} & \textbf{Maturity} \\ \midrule
Flux-qubit & Sparse & Analog & Thermal & Commercial \\ 
LHZ & All-to-all & Analog & Medium & Experimental \\ 
Rydberg & Long-range & Mixed & Low & Prototype \\ \bottomrule
\end{tabularx}
\end{table}

\subsection{Advanced Control Features}

Modern quantum annealers support sophisticated control beyond simple linear annealing schedules. Time-dependent schedules A(s) and B(s) govern the relative weights of transverse-field driver and problem Hamiltonians respectively, where $s_i \in \{-1, +1\}$ parameterizes annealing progress. Non-linear schedules can redistribute evolution time to slow near critical spectral gaps and accelerate through favorable regions\cite{ohkuwa2018,jattana2024,crosson2021}.

Reverse annealing initializes the system in a classical state (rather than quantum superposition), temporarily increases quantum fluctuations to enable local exploration, then re-anneals to refine the solution. This protocol effectively implements quantum-assisted local search, proving particularly valuable in hybrid workflows where classical heuristics generate seed solutions that quantum annealing subsequently refines. Empirical studies demonstrate that reverse annealing can mitigate adverse effects of first-order phase transitions, and when combined with pause protocols (temporarily halting evolution near minimum spectral gaps to exploit thermal relaxation), can substantially improve success probabilities.

\subsection{Error Sources and Mitigation}

Quantum annealing devices face multiple error sources: thermal excitations from finite operating temperature (typically 10-20 mK), decoherence from environmental coupling, analog control imprecision in coefficient realization, inter-qubit crosstalk, and measurement errors in final state readout. These perturbations cause undesired excitations, parameter variations, and misclassification of final spin configurations, degrading both optimization performance and sampling fidelity  \cite{amin2023}. Contemporary error mitigation strategies adapted for quantum annealing include zero-noise extrapolation (extrapolating performance from multiple runs at different noise levels), temperature rescaling (modeling thermal noise effects), and energy-time rescaling to approximate ideal adiabatic behavior [22]. Quantum annealing correction (QAC), discussed in Section 4.4, employs encoding redundancy to suppress logical errors without requiring full fault tolerance. While these techniques improve performance, they do not eliminate fundamental precision and coherence limitations that ultimately bound achievable solution quality\cite{amin2023,pudenz2015,nambu2024}.

\section{QA Algorithmic Strategies}
Quantum annealing (QA) is not a specific algorithm, but is a collection of dynamical methods that are used to direct a quantum system to the low-energy states of a problem Hamiltonian. The quality of exploration of complex energy manifolds of QA is controlled by algorithmic decisions, performance scheduling, a driver code, error-suppression codecs, and the application of hybrid classical-quantum feedback loops.

\subsection{Standard QA: Dynamics and Tunneling}

Standard QA evolves the system under a time-dependent Hamiltonian of the form

\begin{equation}
H(s) = A(s)H_{\text{driver}} + B(s)H_{\text{prob}}, \quad s \in [0,1]
\end{equation}

where $H_{\text{driver}}$ induces quantum fluctuations and $H_{\text{prob}}$ encodes the optimization objective. In the prototypical transverse-field Ising setting, decreasing A(s)and increasing B(s)shifts weight from quantum delocalization to classical energy minimization. QA is often compared to simulated annealing (SA): SA uses thermal fluctuations to hop between basins, while QA leverages quantum tunneling to traverse barriers. In many rugged landscapes, tunneling can cross narrow but tall barriers that stymie thermal walkers, leading to different empirical scaling behaviors between QA and SA on some instance families[10]. In a tunnel-favoring barrier-shape geometry, such as tunnel-favoring distances between two barriers, QA is superior to classical hill-climbing: in an entropic, artificially broad-basin geometry, thermal strategies are better, or hybrid workflows \cite{hauke2020,tasseff2024}.

\subsection{Advanced Schedules}

Nonlinear annealing schedules redistribute time around critical gaps using problem-aware forms of A(s)and B(s), slowing near bottlenecks and accelerating in slack regions.
Adaptive annealing dynamically adjusts schedules based on intermediate signals, offering potential quadratic improvements and reduced effective anneal times.
The pause–relax–resume strategy introduces a pause at $s\ast$ to exploit thermal relaxation and dissipation, empirically improving success rates near diabatic bottlenecks 
 \cite{hauke2020}.

\subsection{Reverse Annealing}

Reverse annealing initializes the system in a classical seed state rather than the uniform superposition required by standard annealing, then increases the transverse field to enable localized quantum exploration before re-annealing. This protocol functions as a quantum local search operator, refining solutions obtained from classical heuristics or previous annealing runs. By avoiding global quantum superposition, reverse annealing can efficiently escape suboptimal basins or correct defects in initial seeds without requiring full global optimization.
Hybrid methods interleaving forward and reverse annealing demonstrate superior performance compared to forward-only protocols across multiple applications including portfolio optimization and matrix factorization. These workflows exemplify the broader principle that quantum annealing often provides greatest value not as a standalone solver, but as a specialized component within classical-quantum computational pipelines.

\subsection{Quantum Annealing Correction (QAC)}

Control errors, thermal fluctuations and decoherence in analog annealers can be prevented by quantum annealing correction (QAC), which suppresses errors instead of corrects them. Modest encoding of logical qubits with repetition or parity imposing penalty Hamiltonians are used; low energy excitations are therefore correctible logical error events. Common QAC protocols include energy penalty couplings, redundancy and majority or parity decoding encodings and classical post-readout decoding. QAC does not provide fault tolerance as does fault-tolerant circuit-based quantum error correction; however, it is highly scalable (near-term), in practice, with a particular improvement to hardware \cite{kadowaki1998,amin2023,pudenz2015,nambu2024}.

\subsection{Non-Stoquastic Drivers}

By satisfying the condition that its off-diagonal terms are real and non-positive in computational basis, a Hamiltonian is a stoquastic which can be simulated by quantum Monte Carlo efficiently by avoiding the sign problem. Introduction of non-stoquastic drivers (e.g., improperly signed $\sigma_i^x\sigma_j^x$couplings) breaks this property, the dynamics acquired being classically hard to simulate. The spectral structure, e.g., first-order phase transitions being reduced to second-order transitions and exponentially small gaps being avoided, and allowing quantum interference and new tunnelling routes previously inaccessible to stoquastic annealing, can all be modified by such terms.\cite{bravyi2015,marvian2019,vinci2017}.

\subsection{Trotterization, QAOA, and Hybrid Strategies}

\begin{table}[htbp]
\centering
\caption{Comparison of Quantum Optimization Paradigms (QA, QAOA, VQE)}
\label{tab:aspect_comparison}
\footnotesize 
\renewcommand{\arraystretch}{1.5} 
\definecolor{lightgrey}{gray}{0.95} 

\rowcolors{2}{lightgrey}{white} 

\begin{tabularx}{\columnwidth}{@{} >{\bfseries}l X X X @{}}
\toprule
\textbf{Aspect} & \textbf{QA} & \textbf{QAOA} & \textbf{VQE} \\ \midrule
Model & Analog & Digital & Variational \\ 
Objective & Sampling & Approximation & Eigenvalue \\ 
Depth & Constant & Increasing & Variable \\ 
NISQ viability & High & Medium & Medium \\ 
Embedding & Required & Not needed & Not needed \\ \bottomrule
\end{tabularx}
\end{table}
Quantum annealing (QA) and the Quantum Approximate Optimization Algorithm (QAOA) could be considered as quantum copies in a consistent quantum control setting where QAOA will simulate transverse-field dynamics over time in the large-p (Trotterised) approximation. On the other hand, digitized annealing and diabatic pulse schedules are similar to optimized finite-depth QAOA, making it easier to hybridize the two modalities to transform their understanding. This structural parallelism has the strength of informing optimal QAOA parameter choices and encourages the use of the Hamilton reordering and Trotterisation methods to transform continuous time annealing into effective gate-based codes \cite{thanh2024,pudenz2015,nambu2024,acharya2025,bravyi2015,marvian2019,tasseff2024}.
\begin{table}[htbp]
\centering
\caption{Taxonomy of Quantum Annealing Algorithmic Variants}
\label{tab:qa_taxonomy}
\scriptsize 
\renewcommand{\arraystretch}{1.6}
\definecolor{rowgrey}{gray}{0.95}

\rowcolors{2}{rowgrey}{white}

\begin{tabularx}{\columnwidth}{@{} >{\bfseries}l X X X X @{}}
\toprule
\textbf{Strategy} & \textbf{Core Idea} & \textbf{Benefit} & \textbf{Cost} & \textbf{Best Use Case} \\ \midrule
Standard QA & Global anneal & Broad exploration & Gap sensitivity & Unknown optima \\ 
Pausing & Thermal relaxation & Higher success & Tuning & Small gaps \\ 
Reverse QA & Local refinement & Fast convergence & Seed needed & Hybrid loops \\ 
QAC & Encoding redundancy & Noise suppression & Qubit overhead & Noisy devices \\ 
Non-stoquastic & Gap reshaping & Potential speedup & Hardware & Hard instances \\ \bottomrule
\end{tabularx}
\end{table}

\section{Embedding and Mapping: THE UNSPOKEN BARRIER}

\subsection{QUBO Encoding Strategies}

The minimization of QUBO and Ising Hamiltonians are an inherent feature of quantum annealers; however, realistic optimization problems involving integer or categorical variables need to be optimized with a binary representation first of all. The encoding scheme selected has a direct impact on the number of qubits needed, the scale of penalty terms, the accuracy of coefficient values, and finally whether a solution can be obtained or not. One-hot encoding is a simple and commonly used method of permutation problems, but it is resource intensive, that is, it requires many variables, high penalty strengths, and high hardware precision. Domain-wall encoding encodes integer values using binary-chain transitions and thus quadratic penalty terms and coupling are minimized in addition to allowing more effective high-throughput optimization \cite{codognet2025a,kumagai2021,kikuchi2025,xavier2024,codognet2025b}.

\begin{table}[htbp]
\centering
\caption{QUBO Encoding Strategies}
\label{tab:qubo_encoding}
\scriptsize 
\renewcommand{\arraystretch}{1.6}
\definecolor{rowgrey}{gray}{0.95}

\rowcolors{2}{rowgrey}{white}

\begin{tabularx}{\columnwidth}{@{} >{\bfseries}l X X X X @{}}
\toprule
\textbf{Encoding} & \textbf{Variables} & \textbf{Constraints} & \textbf{Precision Need} & \textbf{Scalability} \\ \midrule
One-hot & High ($N^2$) & Quadratic & High & Poor \\ 
Domain-wall & Medium ($N \log N$) & Linear & Moderate & Better \\ 
Binary & Low ($\log N$) & Higher-order & Very high & Risky \\ \bottomrule
\end{tabularx}
\end{table}

\subsection{Minor Embedding}

After QUBO formulation, problems must be embedded onto the sparse physical connectivity graph of quantum annealing hardware. Since most optimization problems naturally define dense or complete logical graphs, direct implementation is typically impossible. Minor embedding resolves this by representing each logical variable through a connected subgraph (chain) of physical qubits coupled ferromagnetically  \cite{raymond2020}.

\noindent This process introduces multiple challenges:
\begin{enumerate}[leftmargin=*, nosep, label=\textbf{\arabic*.}]
    \item \textbf{Qubit overhead}: Chains typically require 5--12 physical qubits per logical variable depending on problem structure and hardware topology [33,34,35].
    \item \textbf{Chain breaks}: Thermal noise and control imprecision can cause physical qubits within a chain to assume different values, fragmenting the logical variable and producing invalid solutions.
    \item \textbf{Energy scale reduction}: Ferromagnetic chain couplings consume available coupling strength, reducing the effective energy scale for problem-specific terms.
    \item \textbf{Noise amplification}: Longer chains exhibit greater susceptibility to thermal excitations and control errors, degrading solution fidelity.
\end{enumerate} \cite{raymond2020,pudenz2014,vinci2015}.

\subsection{Mapping Costs and Hardware Overheads}

Massive penalty coefficients induced by one-hot encodings, combined with long chains of qubits, often make use of hardware beyond the precision of the hardware, create noise, and make errors spread out along broken chains that must be post-processed. These are phenomena that compromise the fidelity of solutions to thermal fluctuations, control imprecision and variations between physical qubits that represent a single logical variable. Quantum annealing correction (QAC) deals with them by using repetition encoding and energy penalties, which experimentally increases the probability of successful ground states. QAC also increases performance on NP-hard planted problems when it is used in combination with embedding 
\cite{kumagai2021,xavier2024,raymond2020,pudenz2014,vinci2015}.

\begin{table}[htbp]
\centering
\caption{Logical $\rightarrow$ Physical Qubit Inflation}
\label{tab:qubit_inflation}
\footnotesize 
\renewcommand{\arraystretch}{1.5}
\definecolor{rowgrey}{gray}{0.95}

\rowcolors{2}{rowgrey}{white}

\begin{tabularx}{\columnwidth}{@{} >{\bfseries}l X X l @{}}
\toprule
\textbf{Problem Type} & \textbf{Logical Vars} & \textbf{Physical Qubits} & \textbf{Ratio} \\ \midrule
MaxCut & 100 & 800 & 12.5\% \\ 
TSP (20 cities) & 400 & 3,000+ & $<$10\% \\ 
Portfolio (50 assets) & 50 & 600 & $\sim$8\% \\ \bottomrule
\end{tabularx}
\end{table}

\section{Benchmarks: Critical Analysis}
Quantum annealing is a controversial field, though its perceived results in terms of benchmarking and other tasks remain so. Although there are many articles that underlie better performance, lack of standard operating procedures, other inconsistent backgrounds and inadequate validation of experimental manipulations still present incoherent results. The current part outlines the appropriate benchmarking principles.
\subsection{Benchmarking Principles}

Controlled scaling studies are also easier with synthetic benchmarks like random spin -glass or planted -solution QUBOs, but typically scale poorly with real applications because of their highly idealized structural approximations. More balanced evaluation relies on systematic issues such as the travelling salesman problems, scheduling, knapsack and graph partitioning. To prove practical benefit, it is required to be tested on real-world examples (such as vehicle routing, portfolio optimization, logistics) which are typically addressed using hybrid quantum-classical algorithms and which are classically hard with or without QA. Strict solver parity, i.e. equal preprocessing, accuracy, restarts, and time budgets which are recognized to strongly depend on input/output overhead, model structure factors, and other systemic factors, which when ignored have an exaggerated effect on perceived quantum advantage, is required in rigorous benchmarking  \cite{ignatov2024}.

\subsection{Common Methodological Flaws}

\noindent Weak benchmarking frequently exhibits several characteristic deficiencies:
\begin{enumerate}[leftmargin=*, nosep, label=\textbf{\arabic*.}]
    \item \textbf{Selective reporting}: Presenting only best-case timing or solution quality while omitting failure rates, median performance, and statistical distributions.
    \item \textbf{Inadequate baselines}: Comparing against naive classical implementations (random search, greedy heuristics, unoptimized simulated annealing) rather than industrial-grade solvers (Gurobi, CPLEX, SCIP) or mature metaheuristics (sophisticated genetic algorithms, tabu search with memory structures, hybrid evolutionary strategies).
    \item \textbf{Cherry-picked instances}: Evaluating only small problem sizes that fit hardware constraints without demonstrating scaling to practically relevant dimensions.
    \item \textbf{Omitted overhead}: Reporting only annealing time while excluding encoding formulation, embedding computation (often seconds to minutes), and post-processing for constraint repair.
    \item \textbf{Ignored preprocessing impact}: Failing to account for classical preprocessing (decomposition, variable reduction, backbone identification) that may contribute more to performance than quantum annealing itself \cite{ref37, ref36, ref38}.
\end{enumerate} \cite{ignatov2024,osaba2021,gou2025}.

\begin{table}[htbp]
\centering
\caption{Benchmarking Do's and Don'ts}
\label{tab:benchmarking_practices}
\footnotesize 
\renewcommand{\arraystretch}{1.5}
\definecolor{rowgrey}{gray}{0.95}

\rowcolors{2}{rowgrey}{white} 

\begin{tabularx}{\columnwidth}{@{} l X X @{}}
\toprule
\textbf{Practice} & \textbf{Correct Approach} & \textbf{Common Mistake} \\ \midrule
Runtime & Wall-clock & Anneal time only \\ 
Statistics & Median + variance & Best-case only \\ 
Baselines & Gurobi/Tabu/SA & Naive heuristics \\ 
Instances & Real + synthetic & Toy problems \\ \bottomrule
\end{tabularx}
\end{table}

\subsection{Comparison Against Industrial-Grade Solvers}

Benchmark constructive QA methods should be compared to classical solvers of industrial grade and not heuristics. Such comparisons involve simulated annealing, tabu search, and mixed-integer linear programming solvers (e.g. Gurobi, CPLEX, SCIP) that are sensitive to modelling and constraint-construction mistakes. Contemporary meta -heuristics, such as genetic algorithm, particle swarm optimization, ant-colony optimization, and differential evolution have strong global search algorithms and to this date, no QA problem has been proved to be consistently better than these protocols at solving real-world problems  \cite{tomar2023}.

\subsection{Empirical Evidence Today}

QA can be useful when dealing with rugged energy landscapes that have narrow barriers, where tunnelling can be used; it can also be effective in sampling clustered low-energy basins when penalties are tuned, and can also be effective in hybrid workflows (as in tabu-QA or backbone-QA) compared to going straight to QA the ability to sample rugged energy landscapes by classical decomposition and pruning is in part responsible. On the other hand, QA has difficulty with sparsely connected penalty structures where tunnelling gains are reduced by penalties, integer-intensive formulations where QUBO blow-up and penalty saturation occur and when noise in analogue coefficient derivatives obscures fine scale differences between objectives \cite{gou2025,bonomi2022}.

\section{Application Domains}
The quantum annealing theory has since developed beyond the abstract analysis to have an application-focused optimization context; however, the utility of such techniques as a whole now is still subject to the compatibility of the problem structure, the overhead of problem encoding, and how many such problems can be scaled to using current quantum hardware.

\subsection{Transportation and Logistics}

\subsubsection{Problem Formulation}
Routing problems constitute a preeminent category of transportation optimization tasks, including the travelling salesman problem (TSP), the vehicle routing problem (VRP), the capacitated VRP (CVRP), and a range of trip optimization problems. These are issues of assigning discrete paths to minimize the cumulative travelling cost while meeting capacity, time-window, and service constraints. Since the feasible solution space of routing problems increases factorially with the number of nodes, routing problems are NP-hard \cite{mario2025,chow2025}.

\subsubsection{QUBO Encoding}
Routing problems are mapped using binary decision variables:
\begin{equation}
    x_{i,t} = 
    \begin{cases} 
    1 & \text{if city } i \text{ is visited at time } t \\
    0 & \text{otherwise}
    [cite_start]\end{cases} 
\end{equation}
The objective function is defined as:
\begin{equation}
    [cite_start]\min \sum_{i,j,t} d_{ij} x_{i,t} x_{j,t+1} [cite: 257]
\end{equation}
Each city visited exactly once, one city per time step and Capacity constraints via auxiliary variables This leads to QUBOs with dense coupling matrices and large penalty coefficients.
\subsubsection{Hardware Model}
Clusters Superconducting quantum-annealers, such as the D-Wave system, are commonly used in hybrid classical-QA routing solvers, which use clustering heuristics and forward and reverse annealing regimes to optimize the solutions of the optimization problem.

\subsubsection{Results}
Hybrid QA routing solvers have a time constraint of creating paths of medium-size CVRP under a period of less than a second. With cold-chain VRP, the quality of routes calculated with QA techniques have been seen to be roughly 85 per cent of those calculated with conventional techniques and run-times are roughly 95 per cent shorter than genetic algorithms.

\subsubsection{Limitations}
The erosion of quantum advantage in high-dimensional QUBO programs to encode highly descriptive constraints is exponential, and the number of ancillary variables is rapidly increasing with the number of planning variables, which are measured with physical quantities, in operation, such as time-window feasibility and multi-depot logistics.

\subsection{Energy Optimization (Unit Commitment and Grid Control)}
\subsubsection{Problem Formulation}
Unit commitment problem is the on-off problem of the cycles of power generators with reference to maximize the use of fuel, start-up and ramping costs at the same time keeping balance equations of power and reserve limitations. This optimization is normally formulated using binary, time dependent decision variables[cite: 269].

\subsubsection{QUBO Encoding}
Let $x_{g,t}$ denote the on/off state of generator $g$ at time $t$; the formulation includes linear operating costs, pairwise ramping penalties, and quadratic load-balancing constraints, with higher-order reliability constraints reduced via auxiliary binary variables.

\subsubsection{Hardware Model}
Next-generation scheduling techniques use global forward annealing to do coarse-grained optimization, reverse annealing to do fine-grained ramp refinement, and depend on mandatory hybrid decomposition methods to deal with complexity.

\subsubsection{Limitations}
Variable temporal coupling creates densely connected QUBOs that need penalty preciseness that is higher than the hardware precision capacity; hence, even grid-scale scheduling issues cannot be soluble in the near future.

\subsection{Robotics and Motion Planning}

Motion-planning, task-scheduling, and grasp-selection use binary variables to code path segments and joint configurations \cite{fiqa2025}.
\subsubsection{Problem Formulation}
The main areas of robotic optimization are motion-planning, task-scheduling, grasp-selection and stereo-correspondence.

\subsubsection{QUBO Encoding}
In each of these, binary variables are used to code the choice of path segments, the configurations of arm joints and the sequence of execution of tasks.

Objective:
\begin{equation}
    \min \sum_{i} c_i x_i + \sum_{i<j} J_{ij} x_i x_j
\end{equation}
where $c_i$ represents energy, distance, or execution cost.

\subsubsection{Hardware Model}
Superconducting QA on cloud-based systems is an integrated part of designed robot-QA co-simulation designs.

\subsubsection{Results}
The beneficial results of these integrated control architectures have been optimization of deformable-object manipulation, arm dynamics and stereo-matching using QUBO/Ising encodings.

\subsubsection{Limitations}
When a robot is dynamical, the size of QUBOs grows super-linearly with workspace resolution, and real-time operation of closed-loop control is impossible with existing hardware.

\subsection{Finance: Portfolio Optimization}

Portfolio optimization aims to maximize the expected performance with minimum risk taking into consideration the transaction costs and portfolio rebalancing using binary variables to select the assets and discretized continuous weights \cite{morapakula2025}.

\subsubsection{QUBO Encoding}
Binary asset-selection QUBO:
\begin{equation}
    \min_{i} (-\mu_i) x_i + \sum_{i<j} \Sigma_{ij} x_i x_j
\end{equation}
Transaction costs and cardinality constraints encoded via quadratic penalties.

\subsubsection{Hardware Model}
The hybrid quantum-classical investment process combines classical wealth allocation algorithms with QA-derived asset selection, reverse annealing in refining portfolios, and iterative forward anneals in rebalancing portfolios.

\subsubsection{Results}
Experimental research has shown that hybrid QA-classical approaches are able to generate cumulative returns available to be above the benchmarks of traditional funds, and rebalanced QA portfolios are outperforming unadapted (static) QA portfolios.

\subsubsection{Limitations}
Asset correlations yield dense QUBOs, discretization of weights variables distorts the efficient frontier, and transaction cost model is highly sensitive with regard to the calibration of penalty terms.

\subsection{Drug Discovery and Materials Design}
\subsubsection{Problem Formulation}
Tasks in molecular design such as optimization of molecular structure and reaction pathways as well as design with properties can be re-formulated as discrete graph-labeling problems and as constraint-satisfaction problems.

\subsubsection{QUBO Encoding}
In this formulation, binary variables are used to encode the choice of atoms, bond structures and inclusion of substructures and the objective function provides penalties that apply to property violations, structural soundness, and stabilization issues.

\subsubsection{Hardware Model}
Hybrid quantum-classical machine-learning pipelines are based on an integration of QA-based discrete search and classical property evaluation modules.

\subsubsection{Results}
The hybrid quantum machine learning models improve molecular sub-structure recognition, but with fewer parameters than traditional methods, as well as, the predictive power of sub-structure models.

\subsubsection{Limitations}
Right assessment of molecular validity needs enormous penalty coefficients, which boosts the complexity of QUBO encodings concerning standard physics-simulation solutions; hence any quantum gain is territory-limited to small step frames of structure-selection.

\subsection{Machine Learning}

\subsubsection{Problem Formulation}
Essentially every discrete machine-learning problem, including feature selection, instance selection, and clustering can be formulated as a looking-glass problem of optimization, e.g. redundancy minimization, subset selection and medoid identification \cite{pomeroy2025}.

\subsubsection{QUBO Encoding}
Feature selection QUBO:
\begin{equation}
    \min \sum_{i} w_i x_i + \sum_{i<j} r_{ij} x_i x_j
\end{equation}
where $w_i = $ feature relevance penalty and $r_{ij} = $ redundancy penalty. Clustering uses medoid-selection QUBOs.

\subsubsection{Hardware Model}
The workflow employed makes use of D-wave advantage system and the use of classical preprocessing, QA-based refinement, and post-processing decoding.

\subsubsection{Results}
Quantum annealing can be more feature- and cluster-compact and has better indicators of feature and cluster compactness as well as retrieval (nDCG and Davies Bouldin index) relative to simulated annealing even with natural hardware constraints.

\subsubsection{Limitations}
The effects of feature-correlation result in dense QUBOs, and the data scaling capabilities outrun embedding limits by forcing the use of QA to only refine an objective, and not fully solve it. In routing, robotics, finance, drug discovery, and machine-learning applications, quantum annealing has been most likely to empirically prove useful as a hybrid local optimizer, used on data that has been classical preprocessed to minimize its search space and identify regions dominated by barriers.

\subsection{Application Domain Summary}
\begin{table}[htbp]
\centering
\caption{Application Domain Summary}
\label{tab:domain_summary}
\footnotesize
\renewcommand{\arraystretch}{1.5}
\rowcolors{2}{rowgrey}{white}
\begin{tabularx}{\columnwidth}{@{} l X X X X @{}}
\toprule
\textbf{Domain} & \textbf{Problem} & \textbf{Encoding} & \textbf{Hardware} & \textbf{Outcome} \\ \midrule
Logistics & VRP & One-hot & Flux QA & Hybrid gains \\ 
Energy & Unit commitment & Binary & Hybrid & Classical wins \\ 
Finance & Portfolio & Binary & Reverse QA & Competitive \\ 
ML & Feature selection & QUBO & QA & Compact models \\ \bottomrule
\end{tabularx}
\end{table}

\section{QA vs Other Quantum Paradigms}
Quantum annealing is often introduced as a separate optimization theory; that its practical importance is best understood in terms of its contrasts to the gate-based variational algorithms like the Quantum Approximate Optimization Algorithm (QAOA) and the Variational Quantum Eigen solver (VQE).
\subsection{QA vs QAOA}

The quantum annealing (QA) system provides a continuous-time evolution using a Hamiltonian (where an evolution is performed by switching the transverse-field driver to the problem Hamiltonian). There it is the Quantum Approximate Optimization Algorithm (QAOA) which achieves a similar process in discrete time by Trotterising the evolution of gate-based unitarizes. Shallow QAOA can be considered a discretized approximation of QA where tunable angles are used in place of continuous time, or QA is an attempt to use fixed interactions of a hardware with no construction of an explicit logical circuit. Because of its sampling complexity, shallow effective depth and hardware scalability, QA now perform better on rugged Ising-type optimization problems than the shallow QAOA, but deep QAOA is infeasible on near-term-scale quantum (NISQ) systems  \cite{pelofske2024}.

\subsection{QA vs VQE}

Variational quantum eigensolver (VQE) is simply an approximation method-based eigenvalue solver; a prepared parameterized state is optimized to minimize the expected cost of a Hamiltonian. Optimization is performed in a continuous parameter space, in contrast to quantum annealing, which is the optimization of a discrete energy-space. QA is a natural generator of ensembles of low-energy discrete solutions and the visualizer of combinatorial energy landscapes, which makes it a good model in decision problems. Conversely, VQE applies to a single quantum state when used in physics and chemistry, and the design of ansatz is primarily related to its performance; it is not until it has been heavily discretized that it becomes combinatorial \cite{li2025}.

\section{Current Limitations}

\subsection{Fundamental Limitations}

Embedding overhead represents the most significant practical barrier. Current architectures require 5-12 physical qubits per logical variable, reducing effective problem capacity by 80-92\%. This overhead is fundamentally topological—a consequence of mapping dense logical graphs onto sparse physical connectivity—rather than an engineering deficiency amenable to incremental improvement.
Stoquastic constraints limit the problem classes where quantum advantage is theoretically possible. Standard transverse-field annealers implement stoquastic Hamiltonians that avoid the quantum Monte Carlo sign problem, enabling efficient classical simulation. Consequently, any quantum advantage must arise from practical constant-factor speedups rather than asymptotic complexity separation—a significantly weaker claim than often suggested.
Precision barriers emerge from the interaction of hardware limitations (~100:1 coefficient dynamic range) and problem requirements (often 1000:1 ranges for constraint-heavy formulations). This mismatch causes constraint violations in 30-60\% of samples, necessitating classical post-processing that eliminates supposed quantum advantages.
Benchmarking inconsistencies obscure true performance through methodological flaws: selective reporting, weak baselines, omitted overhead, and inadequate statistics. Until the community adopts rigorous standards, quantum advantage claims remain unsubstantiated \cite{quinton2025,choi2025}.

\subsection{Addressable Research Challenges}

Automated embedding with performance guarantees represents a high-priority research direction. Current heuristic embedding algorithms produce solutions 2-3× worse than theoretical optima and exhibit unpredictable instance-dependent variation. Developing embedding methods with provable approximation ratios or machine learning approaches that predict embedding quality could substantially improve effective hardware utilization.
Non-stoquastic control development offers potential for exponential speedups by avoiding classical simulability constraints. However, implementing stable non-stoquastic evolution requires hardware advances beyond current commercial systems, and empirical evidence for practical advantage remains limited.
Hybrid architecture optimization deserves increased attention. Since practical quantum advantages emerge almost exclusively within hybrid workflows, systematic design principles for classical-quantum problem decomposition, information flow between components, and adaptive resource allocation could maximize utility from current hardware.
Standardized benchmarking protocols including mandatory reporting of full wall-clock timing, median and percentile performance statistics, comparison against industrial-grade classical solvers, and public benchmark instance repositories would dramatically improve research rigor

\subsection{Open Research Questions}

Several fundamental questions remain unresolved:
1.	Annealing hardness characterization: Can we systematically predict which problem instances will benefit from quantum annealing based on structural properties (barrier geometry, gap structure, basin topology) rather than empirical trial-and-error?
2.	Non-stoquastic advantage scaling: Do non-stoquastic Hamiltonians provide scalable performance improvements, or do implementation challenges (decoherence, control complexity) offset theoretical benefits?
3.	Universal quantum advantage metrics: What constitutes a fair, universal metric for quantum advantage in optimization given the multifaceted nature of performance (time-to-solution, approximation quality, sampling diversity, energy efficiency)?
4.	Optimal hybrid architectures: What principles govern effective integration of quantum annealing with classical preprocessing, QAOA-based refinement, and classical post-processing to maximize practical impact?\cite{hauke2020,tasseff2024,ignatov2024,gou2025,bonomi2022}.

\begin{table}[h]
\centering
\caption{Research Priorities}
\small
\begin{tabular}{@{}p{2.2cm}p{1.5cm}p{1.3cm}p{1.5cm}@{}}
\toprule
\textbf{Challenge} & \textbf{Impact} & \textbf{Feasibility} & \textbf{Timeline} \\
\midrule
Embedding automation & High & Medium & 3-5 years \\
Non-stoquastic QA & Very High & Low-Med & 8-12 years \\
Advantage metrics & High & High & 1-2 years \\
Hybrid principles & High & Med-High & 2-4 years \\
\bottomrule
\end{tabular}
\end{table}

\section{Conclusion}

The current overview involves an elaborate discussion of quantum annealing in solving combinatorial optimization, how it is theoretically presented, portrayed through hardware, algorithmic frameworks, encoding methods, benchmarking protocols, architectural applications and future research directions. There are some important conclusions made with respect to the synthesis.  Quantum annealing is a physics-inspired, specialized, heuristic method of optimization that is fundamentally unlike any classical meta-heuristic, as well as any approach grounded on operations of gate implementations of variational quantum algorithms. The characteristic features of it, which are continuous-time analogue evolution, quantum tunnelling between energy levels, and an innate sampling of the distributions of low-energy states, give it certain strengths to problems with compatible structure, especially rugged energy landscapes with narrow, high barriers.  
The overheads of embedding and encoding are known to be the current limits to scalability, and not the number of qubits in the raw. With a physical qubit sparsity typical of modern hardware, minor-embedding costs about 80-92\% of physical qubits, allowing the sparseness to be mitigated. High-density formulations. The formulations of the problems of high constraint density also require more accurate coefficients and are beyond current hardware abilities. As a result, classical overheads, instead of the decoherence or coherence times, are the major impediments along the path to attaining practical quantum advantage. To this end, one can expect decreasing returns as the number of qubits increases without corresponding increases in the strength of connections and precision.  Experimental advantages are found mainly in hybrid models, in which quantum annealing serves as an optimization subroutine after doing extensive computer preprocessing. In ventures of relevance, stand-alone quantum annealing is not often capable of competing with industrial-scale classical solvers over the relevant problem families at real-world scale. Significant applications to logistics, finance, robotics, and machine learning all make use of highly structured hybrid structures which mix classical decomposition, quantum-assisted exploration and classical post-processing.  Strict benchmarking should form the basis of believable performance measurement. Lacking methodological flaws, such as selective reporting, subpar baselines, missing overheads, and inadequate statistical remunerations, effectively overinflates perceived quantum benefits. Scientific rigor cannot be had without the adoption of exhaustive benchmarking protocols which include full wall clock timing, full statistical distributions, comparison to industrial-grade classical benchmarks and publicly available instance libraries.  
The most prospective technical directions to overcome the current limitation include: non deterministic control that has the potential to deliver in exponential speedups with the hardness of classical simulation, though with enormous hardware costs; automated embedding algorithms that make performance guarantees, to better utilize qubits; to combine QA and QAOA in order to take advantage of the complementary properties of analogue and digital quantum control; and systematic characterization of annealing hardness in order to provide principled problem allocation. Quantum annealing cannot be evaluated separately but is a part of the quantum-classical optimization stack, where physics, algorithms and engineering are inseparable. Its final effect will depend on thoughtful incorporation with theoretical developments (hardness characterization, coding strategies), algorithm innovation (hybrid architectures, adaptive protocols) and engineering development (improved connectivity, accuracy, growth of tolerance, etc.).  Even though the question of universal quantum benefit on combinatorial optimization is yet to be resolved, quantum annealing has already demonstrated its scientific worth as a probe of complex energy landscapes and an explanation of quantum dynamics in many-body systems. Its practical role in the optimization of industry is likely to be limited to niche areas where problem structure matches the strengths of quantum annealing with systems and hybrid integration may be successfully managed. The future development must thus be seen as a mixture in viewpoint, looking at the inherent drawbacks with care, but making an attempt at selective improvements of hardware capacity, algorithm complexity and application-based co-architecture.

\section*{Availability of Data and Materials}

No new data were generated or analyzed during this study. All information discussed in this review is based on previously published literature, which has been appropriately cited.

\section*{Competing Interests}

The authors declare that they have no competing financial interests or personal relationships that could have appeared to influence the work reported in this paper.

\section*{Funding}

This research did not receive any specific grant from funding agencies in the public, commercial, or not-for-profit sectors.

\section*{Authors’ Contributions}

Rudraksh Sharma conceived the structure of the review, conducted the comprehensive literature survey, synthesized and analyzed the material, and wrote the main manuscript. Ravi Katukam provided overall guidance on topic selection and conceptual direction, contributed to critical discussions, reviewed and revised the manuscript for intellectual content. Arjun Nagulapally contributed to refining the scope of the review, assisted in structuring key sections, and provided critical feedback during manuscript revision. All authors read and approved the final version of the manuscript.

\section*{Funding}

The authors acknowledge AIONOS for providing a supportive research environment and encouraging interdisciplinary exploration that facilitated the development of this review. The authors also thank colleagues and peers for valuable discussions and feedback that helped refine the scope and clarity of the manuscript. Additionally, the authors acknowledge the developers and maintainers of open-access scientific literature and computational tools that supported this work.

\end{document}